\documentclass[prd,aps,twocolumn,superscriptaddress,preprintnumbers,nofootinbib,floatfix]{revtex4-2}

\def\DLB{\Delta L\!B}
\def\LV{L\!V}
\def\BV{B\!V}

\usepackage{amsmath}
\usepackage{amssymb}
\usepackage{bm}
\usepackage{dcolumn}
\usepackage{graphicx}
\usepackage{slashed}
\usepackage{longtable}
\usepackage{afterpage}

\begin{document}



\title{%
Verification of the tenth-Order QED contribution to the anomalous magnetic moment of the electron from diagrams without fermion  loops
}


\author{Tatsumi Aoyama}
\affiliation{Institute for Solid State Physics, University of Tokyo, Kashiwa, 277-8581 Japan}


\author{Masashi Hayakawa}
\affiliation{Department of Physics, Nagoya University, Nagoya, 464-8602 Japan }
\affiliation{Nishina Center, RIKEN, Wako, 351-0198 Japan }

\author{Akira Hirayama}
\affiliation{Department of Physics, Saitama University, Saitama, 338-8570 Japan }


\author{Makiko Nio}
\email[]{nio@riken.jp}
\affiliation{Nishina Center, RIKEN, Wako, 351-0198 Japan }
\affiliation{Department of Physics, Saitama University, Saitama, 338-8570 Japan }

\begin{abstract}
A discrepancy of approximately 5$\sigma$ exists between the two known results for the tenth-order QED contribution to the anomalous magnetic moment of the electron, calculated from Feynman vertex diagrams without fermion loops. 
To investigate this, we decomposed this contribution into 389 parts based on a self-energy diagram representation, enabling a diagram-by-diagram numerical comparison of the two calculations. No significant discrepancies were found for individual diagrams. 
However, the numerical differences of the 98 diagrams sharing a common structure were not randomly distributed. 
The accumulation of these differences resulted in the 5$\sigma$ discrepancy. 
A recalculation with increased statistics in the Monte Carlo integration was performed for these 98 diagrams.
By replacing the old values with the new ones for these 98 integrals, we have obtained a revised result of $6.800 \pm 0.128$, 
thereby resolving the discrepancy.
\end{abstract}


%

\maketitle

\section{Introduction}
\label{sec:intro}

The anomalous magnetic moment of the electron, $a_e \equiv (g-2)/2$,  has served as a test of quantum electrodynamics (QED) for nearly eight decades. The latest measurements of $a_e$ have achieved a precision of  0.11 ppb \cite{Fan:2022eto}. 
Significant progress has also been made in determining the fine-structure constant  $\alpha$ using atom interferometry with precisions of 0.20 ppb \cite{Parker:2018vye} and 0.081 ppb \cite{Morel:2020dww}. These advancements
have made the comparison between measurements and the theory of $a_e$ one of the most stringent tests of the Standard Model of elementary particles.
Furthermore, improving the precision of $a_e$ in both experiment and theory has emerged as one of the most promising tools on Earth for probing new physics beyond the Standard Model.

The QED contribution to $a_e$ can be obtained by considering the scattering amplitude of an on-shell single electron by an external magnetic field
in the zero-momentum transfer limit.  Since $a_e$ is a dimensionless number, the contributions of other charged leptons appear through their mass ratios relative to the electron. Thus, the QED contribution to $a_e$ can be divided into four terms, each with a different  dependence on the mass ratios:
\begin{align}
a_e(\text{QED}) = A_1 &+ A_2\left( \frac{m_e}{m_\mu} \right)  +A_2\left( \frac{m_e}{m_\tau} \right)   \nonumber \\
                                     &+A_3\left(  \frac{m_e}{m_\mu},\frac{m_e}{m_\tau} \right) .
\end{align}
The mass-independent term $A_1$ is often called universal because it is the same for all charged leptons.
The QED perturbation theory is a powerful and practical tool to calculate these four terms.
The given $2n$th order of the QED perturbation term  is described by Feynman vertex diagrams having  $n$ loops
and the perturbation expansion leads to
\begin{align}
A_1    = A_1^{(2)} \left( \frac{\alpha}{\pi} \right ) 
+ A_1^{(4)}  \left( \frac{\alpha}{\pi} \right ) ^2+  A_1^{(6)}  \left( \frac{\alpha}{\pi} \right ) ^3+\cdots.
\end{align}
The perturbation series shows rapid improvement in precision since $\alpha = 1/137.035\cdots$.
The second-, fourth-, and sixth-order terms of $A_1$ have been obtained as analytically closed forms \cite{Schwinger:1948iu,Petermann:1957,Sommerfield:1958,Laporta:1996mq,Laporta:2017okg}. The sixth-order  term $A_1^{(6)}$ cannot be expressed solely 
in terms of elementary functions, as polylogarithmic functions also appear. The analytic expression of the eighth-order term $A_1^{(8)}$ involves more complex structures, such as one-dimensional integrals over an elliptic function. Although the entire analytic form of $A_1^{(8)}$ has not yet been resolved, 
more than 1,000 digits of the $A_1^{(8)}$ value are known. So, in a practical sense, the eighth-order term is \textit{analytically} known.

To date, the tenth-order term $A_1^{(10)}$ has been calculated only numerically, through two independent efforts.
One effort is by our group, consisting of T. Aoyama, M. Hayakawa, T. Kinoshita, and M. Nio (AHKN) \cite{Kinoshita:2005sm, Aoyama:2008gy,Aoyama:2008hz,Aoyama:2010yt, Aoyama:2010pk,Aoyama:2010zp,Aoyama:2011rm,Aoyama:2011zy,Aoyama:2011dy,Aoyama:2012fc,Aoyama:2012wj,Aoyama:2014sxa,Aoyama:2017uqe,Aoyama:2019ryr}.
The other is by S. Volkov \cite{Volkov:2019phy,  Volkov:2024yzc}.

The $A_1^{(10)}$ term receives contributions from 12,672 Feynman vertex diagrams. 
The 6,318 diagrams involve at least one electron loop, and the remaining 6,354 diagrams have no electron loop.
The former can be divided into 31 gauge-invariant sets.
The results of AHKN for ten of the 31 sets, consisting of the diagrams obtained by inserting vacuum-polarization loops into the second-order vertex diagram, were
confirmed by the independent calculation~\cite{Baikov:2013ula}.
Recently, S. Volkov further independently calculated these 31 sets and confirmed the results obtained by AHKN  \cite{Volkov:2024yzc}.

We call the latter 6,354 tenth-order diagrams without a fermion loop as Set~V following Ref.~\cite{Kinoshita:2005sm}. 
Our latest published result of Set~V is  \cite{Aoyama:2019ryr}
\begin{align} 
 A_1^{(10)} [ \text{Set V}: \text{AHKN}2019]  =     7.668\,(159).
 \label{eq:AHKN2019}
\end{align}
We continued the numerical evaluation of the Set~V integrals, 
focusing particularly on those with larger uncertainties.
At the end of 2020,
our unpublished result of Set~V was
\begin{align}
 A_1^{(10)} [ \text{Set V}: \text{AHKN}2020]  =     7.604\,(140)~.
 \label{eq:AHKN2020}
\end{align}
We refer to this number, Eq.~\eqref{eq:AHKN2020}, and the entire calculation leading to it, as AHKN2020.

S. Volkov's result of  the Set~V contribution is \cite{Volkov:2019phy} 
\begin{align}
A_1^{(10)} [ \text{Set V}: \text{Volkov} 2019\,\text{PRD}] = 6.793\,(90)~.
\label{eq:Volkov2019_PRD}
\end{align}
The datasets provided in the Supplemental Material of Ref.~\cite{Volkov:2019phy} yield a slightly improved result
 \begin{align}
A_1^{(10)} [ \text{Set V}: \text{Volkov} 2019] = 6.824\,(89)~.
\label{eq:Volkov2019}
\end{align}
Hereafter, we refer to the entire calculation leading to Eq.~\eqref{eq:Volkov2019} as Volkov2019.
The difference in the Set~V result between AHKN2020~\eqref{eq:AHKN2020} and Volkov2019~\eqref{eq:Volkov2019} is $+0.780\, (165)$,  
corresponding to a significance of 4.7$\sigma$.
S. Volkov recalculated the Set~V contribution utilizing an off-shell renormalization scheme different from Volkov2019 \cite{Volkov:2024yzc}.
The new result 
\begin{align}
A_1^{(10)} [ \text{Set V} : \text{Volkov} 2024] = 6.857\,(81)
\label{eq:Volkov2024}
\end{align}
is consistent with Volkov2019 \eqref{eq:Volkov2019} and the difference from AHKN2020 \eqref{eq:AHKN2020} still remains at 4.5$\sigma$.
In Ref.~\cite{Volkov:2024yzc}, it is stated that the two results \eqref{eq:Volkov2019_PRD} and \eqref{eq:Volkov2024} are statistically independent and can be combined. This provides the best estimate:
\begin{align}
A_1^{(10)} [ \text{Set V} : \text{Volkov, combined} ] = 6.828\,(60).
\end{align}

In this paper, we performed a diagram-by-diagram comparison between AHKN2020 and Volkov2019 calculations.
We found no apparent discrepancies in any of the diagrams.  Although based on numerical evidence,
we concluded that both AHKN and Volkov correctly formulated the computer programs for the Feynman integrals of Set~V.

However, the numerical differences between the two integrals
are not randomly distributed around zero. 
Volkov's results tend to be smaller than AHKN's, and many tiny differences accumulate, causing the 5$\sigma$ discrepancy.
A hidden bias must exist in the numerical integration of the Feynman integrals. 

We found that the 5$\sigma$ discrepancy arose from one type of 
diagram that contains one second-order self-energy subdiagram.
In the AHKN calculation, the contributions from these diagrams are represented by 98 Feynman integrals.
We reevaluated the 98 integrals using the same numerical integration algorithm, \textsf{VEGAS} \cite{Lepage:1977sw, Lepage:2020tgj}, as in the previous AHKN calculations, 
but with significantly larger statistics.
By replacing the old values with the new ones for these 98 integrals,
we have obtained the new result of Set~V: 
\begin{align}
A_1^{(10)} [ \text{Set V} : \text{AHKN} 2024] = 6.800~(128),
\label{eq:AHKN2024}
\end{align}
which is smaller by $0.803\,(72)$ from \eqref{eq:AHKN2020} and consistent with \eqref{eq:Volkov2019} and with \eqref{eq:Volkov2024}.
Hence, the 5$\sigma$ discrepancy in the tenth-order QED electron anomalous magnetic moment has been resolved.

\section{notation}
\label{sec:notation}

\begin{figure}[t]
\includegraphics[width=8.5cm]{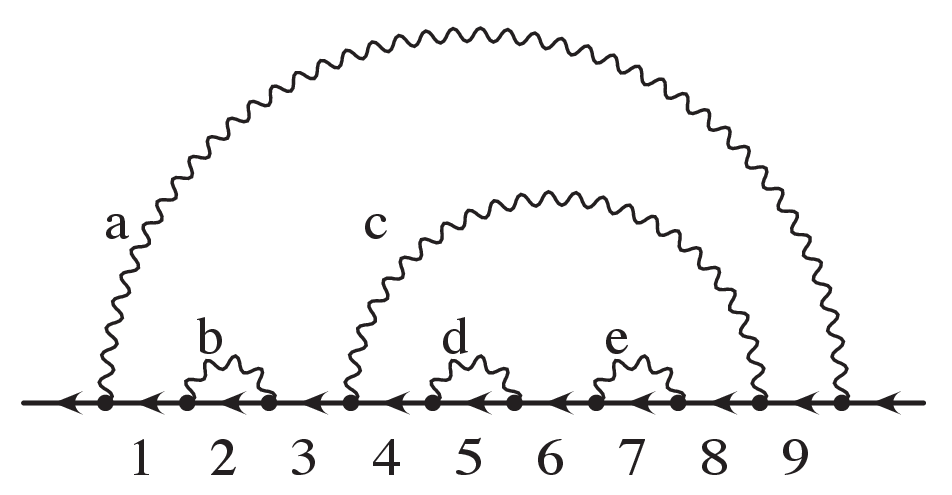}
\caption{
A self-energy diagram of the tenth-order QED.
This diagram is labeled as $X154$ and has a photon index representation
$``abbcddeeca."$  The electron propagators are labeled from 1 to 9 from left to right.
\label{fig:X154}
}
\end{figure}

Let us focus on Feynman diagrams without electron loops.
A one-particle-irreducible self-energy diagram $\mathcal{G}$ of the $2n$th order consists of $n$ photon propagators and $2n-1$ electron propagators. 
In the Feynman gauge for the photon propagator, the $2n$ vertices on the electron line are labeled by the indices of the attached photon lines: $a, b, \cdots, a_n$.
Any self-energy diagram is uniquely identified by the sequence of these vertex labels, read from left to right along the electron line.
See Fig.~\ref{fig:X154} for an example.

A vertex diagram can be obtained by inserting an external photon into one of $2n-1$ electron propagators of the self-energy diagram 
$\mathcal{G}$.  If an external photon is inserted into the $i$th electron propagator, the produced vertex diagram is called $\mathcal{G}(i)$.

The names of self-energy diagrams are determined following previous works \cite{Cvitanovic:1974um,Kinoshita:1981wm,Kinoshita:1990am, Aoyama:2014sxa}.
The naming of diagrams equivalent to their time-reversal counterparts was omitted.
Both the second-order self-energy diagram and vertex diagram are labeled as $2$.
The fourth-order self-energy diagrams are  $4a$ and $4b$ and their photon representations are $``abab"$ and $``abba,"$ respectively.
Eight independent self-energy diagrams of the sixth order exist, labeled from $6a$ to $6h$.
Of the eighth order, there are 47 independent self-energy diagrams, with labels ranging from $01$ to $47$.
The 389 independent tenth-order self-energy diagrams are distinguished by labels from $X001$ to $X389$.

The renormalization constants are named according to the following rules.   The on-shell, AHKN's, and Volkov's vertex renormalization constants derived from a vertex diagram $\mathcal{G}(i)$ are denoted as 
$L_ {\mathcal{G}(i)}$,
$L_ {\mathcal{G}(i)}^\text{UV}$, and  $\LV_ {\mathcal{G}(i)}^\text{UV}$,   respectively.
Similarly, the on-shell, AHKN's, and Volkov's wave-function renormalization constants related to a self-energy diagram $\mathcal{G}$ are denoted as $B_ {\mathcal{G}}$,
$B_ {\mathcal{G}}^\text{UV}$, and  $\BV_ {\mathcal{G}}^\text{UV}$,   respectively.
AHKN and Volkov used the on-shell mass-renormalization constant $dm_{ \mathcal{G}}$.

The bare magnetic moment amplitude obtained from a self-energy 
diagram $\mathcal{G}$ in the AHKN formula is denoted as $M_\mathcal{G}$, 
while that of vertex diagram of $\mathcal{G}(i)$ is denoted as $M_{\mathcal{G}(i)}$.   
The quantity $\Delta M_\mathcal{G}$ is the finite magnetic moment amplitude obtained after subtracting all ultraviolet (UV) and infrared (IR) divergences of   $\mathcal{G}$ using the AHKN recipe.  Similarly,  $\Delta M_{\mathcal{G}(i)}$ is the finite amplitude obtained after subtracting all UV and IR divergences of   $\mathcal{G}(i)$ using Volkov's recipe.  These $\Delta M$ quantities are subject to numerical integration.

In the following, we omit the overall factor $(\alpha/\pi)^n$ of the $2n$th order quantities.
Regularization of UV and IR divergences will be implicit. The regularization parameters are safely removed after the divergent quantities are combined and become finite. The dimensionless quantity $A_1$ does not depend on the explicit value of $m_e$ in any unit system,
and we set $m_e=1$ for simplicity.

 \section{Differences in intermediate procedures}
 \label{sec:IntermediateProcedures}
 
The total contribution of Set~V, consisting of 6,354 vertex diagrams, was calculated by both AHKN and Volkov using numerical integration of Feynman integrals expressed in terms of fourteen Feynman parameters.
However, a diagram-by-diagram comparison of the numerical values of the integrals between AHKN and Volkov is  not straightforward, 
as the intermediate procedures differ between the two.
 
AHKN used the Ward-Takahashi identity to combine all vertex diagrams, which can be obtained by inserting an external photon into an electron propagator of a self-energy diagram. The 6,354 vertex diagrams are then represented by 706 self-energy diagrams.  Since QED is invariant under time reversal, these can be reduced to 389 independent self-energy diagrams: 72 symmetric and 317 asymmetric ones. 
Volkov calculated the contribution of each vertex diagram separately, resulting in the calculation of 3,213 independent vertex diagrams. 

The Ward-Takahashi identity leads to a relation between a self-energy diagram $\mathcal{G}$ and the sum of the vertex diagrams $\mathcal{G}(i)$. For a self-energy diagram $\mathcal{G}$ of the $2n$th order, 
we have 
\begin{align}
\sum_{i=1}^{2n-1} q_\mu \Lambda_{\mathcal{G}(i)}^\mu (p, q) = -\Sigma_{\mathcal{G}}\left(p+\frac{q}{2}\right) + \Sigma_{\mathcal{G}}\left(p-\frac{q}{2}\right) ,
\label{eq:Ward-Takahashi}
\end{align}
where $p\pm q/2$ are the momenta of the electron on the mass shell, and $q$ is the transferred momentum from the external photon.
Differentiating both sides of \eqref{eq:Ward-Takahashi} with respect to $q^\mu$ and neglecting terms quadratic and higher in $q$,
we obtain
\begin{align}
\sum_{i=1}^{2n-1} \Lambda_{\mathcal{G}(i)}^\nu (p, q) \simeq - q_\mu \sum_{i=1}^{2n-1} \left [ \frac{\partial \Lambda_{\mathcal{G}(i)}^\mu (p,q)}{\partial q_\nu} \right ]_{q=0}  
                                                                                     -\frac{\partial \Sigma_{\mathcal{G}}(p)}{\partial p_\nu} ~. 
\label{eq:derivative-Ward-Takahashi}                                                                          
\end{align}
AHKN extracted a contribution to $a_e$ from the right-hand side (rhs) of \eqref{eq:derivative-Ward-Takahashi}, 
while Volkov calculated each vertex diagram $\Lambda^\nu_{\mathcal{G}(i)}$ on the left-hand side (lhs). 

Another difference between AHKN and Volkov is the choice of an intermediate renormalization scheme.   
The on-shell renormalization scheme must be applied in QED calculations to ensure that the coupling constant $e$ and the electron mass $m_e$ are both physical quantities.  However, the renormalization constants defined with the on-shell renormalization scheme suffer from IR divergence. If these constants were used in a numerical calculation to cancel UV divergence, they would introduce new types of IR divergence into the calculation. 
Both AHKN and Volkov circumvent this issue by employing renormalization constants free from IR divergence in their numerical calculations. 

The vertex and wave-function renormalization constants used by AHKN are determined by applying the $\mathbb{K}$ operation 
to vertex and self-energy subdiagrams, respectively \cite{Cvitanovic:1974sv, Aoyama:2005kf}. This is a power-counting rule, and the derived renormalization constants do not satisfy the Ward-Takahashi identity
\begin{align}
 B_{\mathcal{G}}^\text{UV} + \sum_i  L_{\mathcal{G}(i)}^\text{UV} \neq  0~.
 \label{eq:AHKNWT_BL}
 \end{align}
As a result, the residual and finite renormalization must be performed at the end of the calculation to recover gauge invariance. 

Volkov chose the renormalization constants such that they satisfy the Ward-Takahashi identity
\begin{align}
 \BV_{\mathcal{G}}^\text{UV} + \sum_i  \LV_{\mathcal{G}(i)}^\text{UV} = 0~,
 \label{eq:VolkovWT_BL}
 \end{align}
and no residual renormalization is required.

For mass renormalization, both AHKN and Volkov used the on-shell mass-renormalization
constants.  We can therefore neglect mass renormalization in comparing the two
from the outset.

Magnetic moment amplitudes $M_\mathcal{G}$ and $M_{\mathcal{G}(i)}$ may contain IR divergences.
To cancel these IR divergences, AHKN and Volkov introduced IR counterterms consisting of the residual vertex renormalization constants
\begin{align}
L_{\mathcal{G}(i)}^\text{R} \equiv & \quad L_{\mathcal{G}(i)} - L_{\mathcal{G}(i)}^\text{UV}, \nonumber \\
\LV_{\mathcal{G}(i)}^\text{R} \equiv & \quad L_{\mathcal{G}(i)} - \LV_{\mathcal{G}(i)}^\text{UV},
\label{eq:L_IR}
\end{align}
respectively,
where 
$ L_{\mathcal{G}(i)}$ is the on-shell vertex renormalization constant.
These choices of IR counterterms \eqref{eq:L_IR} enable both AHKN and Volkov to realize the on-shell renormalization.
The UV subdivergences in $\LV^\text{R}$ are removed  using the renormalization constants $\LV^\text{UV},  \BV^\text{UV}$,
and $dm$.  For $L^\text{R}$ of AHKN, the UV subdivergences are removed  by the $\mathbb{K}$ operation,
which is equivalent to use  $L^\text{UV},  B^\text{UV}$, and $dm^\text{UV}$.
The mass-renormalization constant $dm^\text{UV}$  differs from the on-shell mass-renormalization constant $dm$
when the order of a subdiagram is the fourth or higher.  Thus, 
the finite mass-renormalization constant $\Delta dm_\mathcal{G}$  along with
the finite vertex constant $\Delta L_{\mathcal{G^\prime }(i) j^\ast}$,
where $j^\ast$ indicates that a two-point vertex is inserted into the $j$th electron propagator of the diagram $\mathcal{G^\prime}(i)$,
appears in the AHKN formulas. For example,  the residual renormalization formula for Set~V, as shown in Eq.~(35) of Ref.~\cite{Aoyama:2014sxa},
includes terms of the form  $\Delta dm_\mathcal{G}\times \Delta L_{\mathcal{G^\prime }(i) j^\ast}$.

AHKN and Volkov used these intermediate renormalization constants to remove UV and IR divergences in bare magnetic
moment amplitudes, rendering them  finite  as $\Delta M_\mathcal{G}$ and $\Delta M_{\mathcal{G}(i)}$,
respectively.
Because of the differences in the intermediate procedures,  AHKN's $\Delta M_\mathcal{G}$ and the Ward-Takahashi-related sum of Volkov's 
$\sum_i \Delta M_{\mathcal{G}(i)} $ inevitably yield different values.
For the tenth-order self-energy diagram $X001$, represented by $``abacbdcede,"$ the integral of
AHKN and the sum of nine integrals of Volkov are
\begin{align}
 \Delta M_{X001} &=  -0.160\,83 \,(334) ,    \nonumber \\
\sum_{i=1}^9  \Delta M_{X001(i)}& =   ~~ 0.580\,95\, (534)~,
\label{eq:DMandDMvX001}
\end{align}
respectively.
In the following,   the numerical values of $\Delta M_{\mathcal{G}}$ and $\Delta M_{\mathcal{G}(i)}$ are taken from the
datasets of integrals that produced 
AHKN2020~\eqref{eq:AHKN2020}
and Volkov2019~\eqref{eq:Volkov2019}, respectively.

\section{comparison method}
\label{sec:ComparisonMethod}

The finite magnetic moments $\Delta M_\mathcal{G}$  and $\Delta M_{\mathcal{G}(i)}$ are derived from the
bare magnetic moment amplitudes subtracting all UV and IR divergences.
We can schematically rewrite the Ward-Takahashi relation of the bare magnetic moment amplitudes \eqref{eq:derivative-Ward-Takahashi}  using the finite magnetic moments as 
\begin{align} 
  0 &=  M_\mathcal{G}   - \sum_i M_{\mathcal{G}(i)}  \nonumber \\
     &= ( \Delta M_\mathcal{G}  + \text{subtraction terms for } M_\mathcal{G}  ) \nonumber \\
     & \hspace{2em} - 
          ( \sum_i {\Delta M_{\mathcal{G}(i)}}   + \text{subtraction terms for}  \sum_i M_{\mathcal{G}(i)}) .
\end{align}
Thus, the difference $\Delta M_\mathcal{G} - \sum_i \Delta M_{\mathcal{G}(i)}$ can be
expressed using the lower-order finite quantities, which are derived from the renormalization constants and magnetic moments.
We will call the symbolic expression of $\Delta M_\mathcal{G} - \sum_i \Delta M_{\mathcal{G}(i)}$  in terms of 
the lower-order quantities as \textit{the gap equation} of $\mathcal{G}$.

\section{difference of renormalization constants $\delta L_{\mathcal{G}(i)}$ }
\label{sec:DiffofRenormalizationConstants}

It is clear that the difference between $\Delta M_\mathcal{G}$ and $\Delta M_{\mathcal{G}(i)}$ can be attributed to 
the difference in the renormalization constants. We  introduce a new constant related to a vertex diagram $\mathcal{G}(i)$ as
\begin{align}
\delta L_{\mathcal{G}(i)} \equiv \LV_{\mathcal{G}(i)}^{\text{UV}} - L_{\mathcal{G}(i)}^{\text{UV}} ~.
\label{eq:deltaL}
\end{align}
The wave-function renormalization constants, $B_\mathcal{G}^\text{UV}$ and $\BV_\mathcal{G}^\text{UV} $, can be converted to the sum of the vertex renormalization constants 
via the Ward-Takahashi relations, \eqref{eq:AHKNWT_BL} and \eqref{eq:VolkovWT_BL}. 
So, only $\delta L_{\mathcal{G}(i)} $ is needed to be known.

$\delta L_{\mathcal{G}(i)}$ can be numerically calculated as follows.
Suppose that $\Lambda^\nu_{\mathcal{G}(i)}(p,q) $ is the proper vertex function of a vertex diagram $\mathcal{G}(i)$.
Then, the  on-shell vertex renormalization constant $L_{\mathcal{G}(i)}$  is obtained
by applying the projection operator to $\Lambda^\nu$ as
\begin{align}
L_{\mathcal{G}(i) } = \frac{1}{4} \mathrm{Tr} [ ( \slashed{p}+1 ) p_\nu \Lambda^\nu_{\mathcal{G}(i)} (p,0) ] ~
\end{align}
with $p^2=1$.

Similarly,  Volkov's vertex renormalization constant $\LV^\text{UV}_{\mathcal{G}(i)}$ defined in Ref.~\cite{Volkov:2019phy}
can be obtained 
\begin{align}
\LV^\text{UV}_{\mathcal{G}(i) } = \frac{1}{12} \mathrm{Tr} [ (  \gamma_\nu -\slashed{p} p_\nu ) \Lambda^\nu_{\mathcal{G}(i)} (p,0) ] ~
\end{align}
with $p^2=1$.

In the Feynman parametric representation, Volkov's vertex renormalization constant of the $2n$th order
has the form
\begin{align}
\LV_{\mathcal{G}(i)}^{\text{UV}}& =  \left(  \frac{1}{4} \right )^n  (n-1)! 
\nonumber \\
&\times \int \frac{(dz)_\mathcal{G} \,z_i} {U^2}
                                    \left [    \frac{F_0}{V^n}  + \frac{F_1}{UV^{n-1} }  + \cdots + \frac{F_n}{U^n V^0}  \right ] ~,
\label{eq:LV_Feynman_parameters}                                        
\end{align}       
where the Feynman parameters $z_j$ are assigned to  internal lines of a self-energy diagram $\mathcal{G}$ 
and are subject to the constraint $(dz)_\mathcal{G} = (\Pi_j dz_j ) \delta(1-z_1-\cdots -z_{2n-1}-z_a-z_b-\cdots-z_{a_n} )$.
The $U$, $V$, and $F_k$ are functions of $z_j$. Their precise definitions are found in, for instance, Refs.~\cite{Cvitanovic:1974uf,Kinoshita:1990am,Aoyama:2005kf}.
The overall logarithmic UV divergence of the vertex diagram $\mathcal{G}(i)$ is contained in the integral involving the $F_n$ term. 
Note that $V^0$ is not $V$ to the zeroth power, but a symbolic expression depending on the UV regularization 
method employed.

The AHKN vertex renormalization constant $L_{\mathcal{G}(i)}^{\text{UV}}$ of the same vertex diagram is determined 
by applying the $\mathbb{K}$ operation to $L_{\mathcal{G}(i)}$ and it exactly corresponds to the integral of the $F_n$ term of \eqref{eq:LV_Feynman_parameters}.
The integral expression of $\delta L_{\mathcal{G}(i)}$ in terms of Feynman parameters is then obvious and well defined.

The new constant $\delta L_{\mathcal{G}(i)}$          
is free from IR divergence by definition of \eqref{eq:deltaL}.
However, it may suffer from UV divergences arising from subdiagrams in $\mathcal{G}(i)$.
To make $\delta L_{\mathcal{G}(i)}$ numerically calculable,  we removed these UV divergences by 
applying the $\mathbb{K}$ operations according to the forest formula \cite{Zavyalov:1965, Zimmermann:1969}.
For a self-energy subdiagram of the fourth order or higher, we also performed the residual mass renormalization, the $\mathbb{R}$ subtraction 
defined in Ref.~\cite{Aoyama:2007bs},
so that the on-shell mass renormalization is realized.
The residual mass-renormalization part being neglected, the finite constant $\delta L_{\mathcal{G}(i)} $ is formally expressed as
\begin{align}
\delta L_{\mathcal{G}(i)} =  \left (1+ \sum_f \left [ \prod_{S_j \in f } ( - \mathbb{K}_{S_j} ) \right ]  \right )
(\LV_{\mathcal{G}(i)}^{\text{UV}} - L_{\mathcal{G}(i)}^{\text{UV}} )~,
\label{eq:forest_fomular_for_dL}
\end{align}
where $f$ is a forest consisting of divergent subdiagrams, each denoted as $S_j$ with different $j$.
According to the forest formula \eqref{eq:forest_fomular_for_dL}, 
both the symbolic expression of the divergent structures 
and the integrand ready for numerical integration can be constructed.

Let us  show the eighth-order vertex diagram $\mathcal{G}(i)=01(1)$, where $\mathcal{G}$ is expressed as  $``abacbdcd,"$ as
an example.
The forest formula and the $\mathbb{K}$ operation  enable us to write down
\begin{align}
\delta L_{01(1)}& =( \LV_{01(1)}^\text{UV} - L_{01(1)}^\text{UV} )  \nonumber \\
&+ \{ - L_{6f(1)}^\text{UV} 
   + 2 L_{4a(1)}^\text{UV} L_2^\text{UV} 
    -( L_2^\text{UV})^3  \} (\LV_2^\text{UV} - L_2^\text{UV} )  \nonumber \\
&
+\{- L_{4a(1)}^\text{UV}    + (L_2^\text{UV} )^2  \}( \LV_{4a(1)}^\text{UV} - L_{4a(1)}^\text{UV} )   \nonumber \\
&
- L_2^\text{UV} ( \LV_{6f(1)}^\text{UV} - L_{6f(1)}^\text{UV} )  ~,                                
 \end{align}
which can be rewritten as a more compact form using the lower-order finite constants:
\begin{align}
\delta L_{01(1)}& = ( \LV_{01(1)}^\text{UV} - L_{01(1)}^\text{UV} )  \nonumber \\
&-  L_{6f(1)}^\text{UV} \delta L_2 
-  L_{4a(1)}^\text{UV} \delta L_{4a(1)}
- L_2^\text{UV} \delta L_{6f(1)}.
\end{align}
The numerical value of $\delta L_{01(1)}$ is obtained by performing ten-dimensional numerical integration with \textsf{VEGAS}:
\begin{align}
\delta L_{01(1)} =  0.660\,67\, (18) \,.
\end{align}
The integral converges well, requiring approximately 30 minutes when computed on an Intel Xeon processor equipped with 40 cores.

Similarly, we obtained the symbolic expressions and numerical values of $\delta L_{\mathcal{G}(i)}$
for all independent vertex diagrams: 1 second-order, 4 fourth-order, 28 sixth-order, and 269 eighth-order diagrams.

\section{diagram-by-diagram comparison using the gap equation}
\label{sec:GapEquation}

To derive the explicit form of the gap equation for  $\Delta M_\mathcal{G} - \sum_i\Delta M_{\mathcal{G}(i)}$,
it is essential to understand the UV and IR divergence structures of $M_\mathcal{G}$ as well as those of $M_{\mathcal{G}(i)}$.
The 389 self-energy diagrams were analyzed by the code-generator program \textsf{gencode}\textit{N} in AHKN's previous work.
The structures of the 3,213 vertex diagrams of Set~V are newly obtained in this work.

The gap equations  are expressed 
in terms of the finite constants $\delta L_{\mathcal{G}(i)}^{(2n)}$   and the finite magnetic moment amplitudes
$\Delta M_{\mathcal{G}}^{(2n)}$  for  $n=1,2,3,4$.  The residual renormalization constants $\Delta L\!B_\mathcal{G}^{(2n)}$  in the AHKN formula 
appear in the gap equation of a diagram containing a self-energy subdiagram of the fourth order or higher. 
For 31 of the 389 diagrams, the finite constants $\Delta dm_\mathcal{G} $ and $\Delta L_{\mathcal{G}(i)j^\ast}$ also appear.
The symbolic expressions and numerical values of  $\Delta M_\mathcal{G}$,  $\DLB_\mathcal{G}$, $\Delta dm_\mathcal{G} $, and $\Delta L_{\mathcal{G}(i)j^\ast}$ were obtained in previous works of AHKN\cite{Aoyama:2014sxa}.  
These lower-order quantities are highly reliable and accurate, as many of them are used to produce the numerical results of $A_1^{(2n)}$ for $n = 1, 2, 3, 4$ of $a_e$, which are consistent with the corresponding analytic results.
 
The gap equation of the diagram $\mathcal{G} = X001$, for instance, is
obtained as
\begin{align}
\Delta &M_{X001} - \sum_{i=1}^9 \Delta M_{X001(i)}  \nonumber \\
  &= \Delta M_2 \,  \{  - 3\,(\delta L_{4a(1)})^2 - 6\,\delta L_{2}\,\delta L_{6f(1)} \nonumber \\
  & \quad + 12\,(\delta L_{2})^2\,\delta L_{4a(1)} - 5 (\delta L_{2})^4 + 2\,\delta L_{01(1)}   \} \nonumber \\
   &+ \Delta M_{4a} \, \{  2\,\delta L_{6f(1)} - 6\,\delta L_{2}\,\delta L_{4a(1)}  + 4\,(\delta L_{2})^3  \}  \nonumber \\
   & + \Delta M_{6f} \, \{ 2\,\delta L_{4a(1)} - 3\,(\delta L_{2})^2 \} \nonumber \\
    & + \Delta M_{01} \, ( 2\,\delta L_{2}) ~.
\label{eq:X001}
\end{align}
The gap equation of the diagram $\mathcal{G}=X154$ shown in Fig.~\ref{fig:X154}, involving a sixth-order self-energy subdiagram $6a$,
is given as 
\begin{align}
 \Delta M&_{X154} - \sum_{i=1}^9 \Delta M_{X154(i)}
\nonumber \\  
     =&  - \Delta M_2 \Delta dm_{6a}  \Delta L_{2^\ast} 
\nonumber \\  
     &  + \Delta M_2 \{  - \delta L_{2}\delta L_{6a3} - 2\delta L_{2}\delta L_{6a(2)} - 2\delta L_{2}\delta L_{6a(1)} 
\nonumber \\     
     & \quad + 2(\delta L_{2})^2  \delta L_{4b(1)} + \DLB_{6a}\delta L_{2} - \delta L_{30(1)} 
\nonumber \\     
      & \quad - \delta L_{18(6)} - \delta L_{18(4)} \}  
\nonumber \\  
     &   + \Delta M_{4b}(  - \delta L_{6a(3)} - 2\delta L_{6a(2)} - 2\delta L_{6a(1)} 
\nonumber \\  
     &   \quad +2 \delta L_{2}\delta L_{4b(1)} + \DLB_{6a} )
 \nonumber \\    
     &   + \Delta M_{6a} (  - \delta L_{4b(1)} )   ~.
\label{eq:X154}
\end{align}
 If Volkov2024 \eqref{eq:Volkov2024} is used instead of Volkov2019 \eqref{eq:Volkov2019}, the
 symbolic expressions of the gap equations remain unchanged. However, the values of $\delta L_{\mathcal{G}(i)}$
 need to be recalculated.
 
We are ready to compare the numerical values of the lhs and the rhs of  the gap equations
for all 389 diagrams.
With \eqref{eq:DMandDMvX001} and the values of lower-order quantities,
 the lhs and the rhs of Eq.~\eqref{eq:X001} are 
$-0.7418\,(63)$ and 
$ -0.7385$,
respectively.
The uncertainty of the rhs is less than 0.0010, as the values of the lower-order quantities in the rhs are well known, with uncertainties less than 0.0002. Therefore, the uncertainty in the rhs is negligible compared to that of the lhs.
The difference  $-0.0033\,(63)$, which is consistently zero,
strongly suggests that both AHKN and Volkov properly created their integrals,
and correctly performed numerical integration for the diagram $X001$.
For the other 388 diagrams, the results of the consistency check are listed in Table~\ref{table:X001}. 

All 389 differences listed in the fifth column of Table~\ref{table:X001} add up to  $+0.782\,(166)$, 
which matches to 
the difference between AHKN2020~\eqref{eq:AHKN2020} and Volkov2019~\eqref{eq:Volkov2019}, $+0.780\,(165)$, as expected.
This happens because the numerical values of the gap equations are reliable and very accurate.

As shown in Table~\ref{table:X001}, there is no apparent inconsistency in any of the 389 diagrams.
We conclude that both the calculation methods employed by AHKN and Volkov, as well as the integrals constructed by the two, are valid.
The observed numerical discrepancy of Set~V is likely due to unknown biases in the numerical integration.

\section{numerical Integration}
\label{sec:Integration}

This section provides a detailed explanation of our new result Eq.~\eqref{eq:AHKN2024}.
To investigate the numerical integration more precisely, the 389 diagrams were divided into several groups based on their self-energy subdiagrams. 

When a self-energy subdiagram in a vertex diagram behaves as a self-mass, the residual diagram becomes the vertex diagram with 
a two-point vertex. The IR divergence of such a diagram is more severe than that of a normal vertex diagram and is inversely proportional to the fictitious photon mass. This linear IR divergence is subtracted using the $\mathbb{R}$ subtraction by AHKN, while it is done using on-shell mass renormalization by Volkov.

The other mechanism of IR divergence is caused when a subdiagram provides a lower-order magnetic moment, and the residual diagram behaves as a vertex diagram. The IR divergence is logarithmic and is handled with the counter term involving Eq.~\eqref{eq:L_IR}. More complex IR divergence is a combination of multiple pieces of the two mechanisms \cite{Aoyama:2007bs}. 

Numerical cancellation of linear or more severe IR divergence yields a steep form of the integrand near the singular region.
As the number of self-energy subdiagrams within a diagram increases, the difficulty of numerically evaluating that diagram also rises.
This is why we paid attention to self-energy subdiagrams.

We found that the difference between AHKN2020~\eqref{eq:AHKN2020} and Volkov2019 \eqref{eq:Volkov2019}
originated almost entirely from one special group: the 98 diagrams containing a second-order self-energy subdiagram and no other self-energy subdiagrams.
The differences listed in the fifth column of Table~\ref{table:X001} for these 98 diagrams add up to $+0.994\,(73)$, accounting for
the 5$\sigma$ discrepancy.

We focused on the 98 diagrams and initiated a new numerical evaluation of their finite magnetic moments, $\Delta M_{\mathcal{G}}$.  All 98 integrals are defined within a twelve-dimensional unit hypercube. The Jacobian of a change of variables from fourteen Feynman parameters to twelve \textsf{VEGAS} variables served as a density function of  Monte-Carlo integration, which was determined based on the structure of a diagram. 
The \textsf{VEGAS} variables were also \textit{stretched} to accelerate the convergence of the integrals \cite{Kinoshita:1990am}.
For the new evaluation, the density functions remained the same as in the previous one, but the stretching parameters were slightly modified.

The number of sampling points of \textsf{VEGAS} integration routine is  $1\times 10^{10} \sim 4 \times 10^{10}$ depending on the complexity of a diagram.
At each sampling point, the integrand is evaluated using double-double-precision real numbers to avoid round-off errors.
Approximately $3.2 \times 10^7 $ core-hours were dedicated to the computations over the two different computer systems.
The new values of the integrals are shown in Table~\ref{table:XB1B2} alongside the previous values used for AHKN2020~\eqref{eq:AHKN2020}.

As shown in Table~\ref{table:XB1B2}, the old and new values of the 98 integrals are consistent with each other.
However, 85 of the 98 integrals decrease in the new calculation
resulting in a shift $-0.8032\,(717)$ for the sum of the 98 integrals.

Each of the 98 integrals contains only one self-energy subdiagram, which allows for relatively good convergence in numerical integration compared to other diagrams with two or more self-energy subdiagrams. Consequently, these 98 integrals had not been reevaluated with increased sampling points exceeding $1 \times 10^{10}$ using double-double precision for real numbers until they were revisited in this work.

The numerical values of the lower-order quantities used in this work are provided as Supplemental Material accompanying this paper.

\begin{acknowledgments}
%
We deeply acknowledge the late Professor Toichiro Kinoshita, whose research on the QED lepton $g\! - \! 2$ laid the foundation for this work.
The numerical works in this paper were conducted on RIKEN's supercomputer systems, HOKUSAI Big Waterfall and  Big Waterfall~2. 
M.~H. is partially supported by JSPS Grant-in-Aid for Scientific Research (C)20K03926.
M.~N. is partially supported by JSPS Grant-in-Aid for Scientific Research (C)16K05338,  (C)22K03646, and (S)20H05646.
\end{acknowledgments}


\bibliographystyle{apsrev}
\bibliography{b}

\afterpage{
\begingroup
\squeezetable
\renewcommand{\arraystretch}{0.8}
\setlength\LTleft{0pt}   
\setlength\LTright{0pt}  
\setlength\LTcapwidth{\textwidth} 
\begin{longtable*}{l@{\hskip-9em}l@{\hskip-5em}d@{\hskip-1em}d@{\hskip-1em}d}
\caption*{%
\small TABLE~\ref{table:X001}:
Consistency check between AHKN2020 and Volkov2019 for the tenth-order
magnetic moment X001--X389 of  the Set~V diagrams.
The first and second columns show the diagram label and 
its representation by photon indices, respectively. 
The third column lists the numerical difference between the integrals of AHKN2020, $\Delta M_{\mathcal{G}}$, and
the sum of the integrals of Volkov2019, $\sum_i \Delta M_{\mathcal{G}(i)}$.
The  fourth column presents numerical values of the gap equations, calculated using lower-order quantities: $\delta L_{\mathcal{G}(i)}$, $\Delta M_{\mathcal{G}}$, $\Delta L\!B_{\mathcal{G}}$, $\Delta dm_\mathcal{G}$, and $\Delta L_{\mathcal{G}(i)j^\ast}$.
The fifth column shows the differences between values in the third and fourth columns. The uncertainties in the fifth column arise entirely from the uncertainties in the tenth-order magnetic moments listed in the third column.
\label{table:X001}
} \\ 
\hline \hline
\multicolumn{1}{p{0.12\textwidth}}{Diagram $\mathcal{G}$} &
\multicolumn{1}{p{0.12\textwidth}}{Expression} &
\multicolumn{1}{p{0.20\textwidth}}{\hspace{2em}$\Delta M_\mathcal{G} - \sum_i \Delta M_{\mathcal{G}(i)}$ }  & 
\multicolumn{1}{p{0.18\textwidth}}{\hspace{3em}Gap Equation} &
\multicolumn{1}{p{0.18\textwidth}}{\hspace{4em}Difference} 
\\
\hline
\endfirsthead
\caption*{%
\small TABLE~\ref{table:X001}(continued):
Consistency check between AHKN2020 and Volkov2019 for the tenth-order
magnetic moment X001--X389 of  the Set~V diagrams.
} \\
\hline \hline
\multicolumn{1}{p{0.12\textwidth}}{Diagram $\mathcal{G}$} &
\multicolumn{1}{p{0.12\textwidth}}{Expression} &
\multicolumn{1}{p{0.20\textwidth}}{\hspace{2em}$\Delta M_\mathcal{G} - \sum_i \Delta M_{\mathcal{G}(i)}$ } & 
\multicolumn{1}{p{0.18\textwidth}}{\hspace{3em}Gap Equation} &
\multicolumn{1}{p{0.18\textwidth}}{\hspace{4em}Difference} 
\\
\hline
\endhead
\hline
\endfoot
\hline \hline
\endlastfoot
$X$001 & $abacbdcede$ &   -0.7418 &   -0.7385 &   -0.0033 \,( 63) \\
$X$002 & $abaccddebe$ &    8.0130 &    8.0253 &   -0.0123 \,(139) \\
$X$003 & $abacdbcede$ &    2.0226 &    2.0221 &    0.0006 \,( 29) \\
$X$004 & $abacdcdebe$ &   -6.5041 &   -6.5146 &    0.0104 \,(130) \\
$X$005 & $abacddbece$ &   -0.2680 &   -0.2789 &    0.0110 \,(106) \\
$X$006 & $abacddcebe$ &    0.5522 &    0.5522 &   -0.0000 \,(125) \\
$X$007 & $abbcadceed$ &   -0.2365 &   -0.2250 &   -0.0115 \,(128) \\
$X$008 & $abbccddeea$ &   -3.8164 &   -3.8115 &   -0.0050 \,(168) \\
$X$009 & $abbcdaceed$ &    0.1962 &    0.2050 &   -0.0089 \,( 69) \\
$X$010 & $abbcdcdeea$ &   -1.4020 &   -1.4014 &   -0.0007 \,(137) \\
$X$011 & $abbcddaeec$ &   -0.6609 &   -0.6645 &    0.0035 \,(110) \\
$X$012 & $abbcddceea$ &   -0.4561 &   -0.4717 &    0.0156 \,(129) \\
$X$013 & $abcabdecde$ &    1.7891 &    1.7879 &    0.0012 \,( 14) \\
$X$014 & $abcacdedbe$ &    0.2015 &    0.2018 &   -0.0003 \,( 32) \\
$X$015 & $abcadbecde$ &    1.1125 &    1.1141 &   -0.0016 \,(  6) \\
$X$016 & $abcadcedbe$ &   -0.4574 &   -0.4567 &   -0.0007 \,(  5) \\
$X$017 & $abcaddebce$ &   -0.7935 &   -0.7966 &    0.0030 \,( 15) \\
$X$018 & $abcaddecbe$ &   -0.4154 &   -0.4194 &    0.0040 \,( 17) \\
$X$019 & $abcbadeced$ &    2.3984 &    2.3968 &    0.0016 \,( 31) \\
$X$020 & $abcbcdedea$ &    7.6668 &    7.6799 &   -0.0131 \,(131) \\
$X$021 & $abcbdaeced$ &    0.3403 &    0.3408 &   -0.0005 \,( 16) \\
$X$022 & $abcbdcedea$ &   -1.1144 &   -1.1238 &    0.0094 \,(114) \\
$X$023 & $abcbddeaec$ &   -0.4987 &   -0.5074 &    0.0087 \,( 58) \\
$X$024 & $abcbddecea$ &    4.5369 &    4.5595 &   -0.0226 \,(131) \\
$X$025 & $abccadeebd$ &    2.4091 &    2.4333 &   -0.0242 \,(119) \\
$X$026 & $abccbdeeda$ &    1.2078 &    1.2042 &    0.0035 \,(110) \\
$X$027 & $abccdaeebd$ &   -0.2238 &   -0.2272 &    0.0034 \,( 46) \\
$X$028 & $abccdbeeda$ &   -4.8204 &   -4.8306 &    0.0101 \,(129) \\
$X$029 & $abccddeeab$ &    3.1830 &    3.1830 &   -0.0000 \,( 95) \\
$X$030 & $abccddeeba$ &   -0.4286 &   -0.4201 &   -0.0085 \,(131) \\
$X$031 & $abcdaebcde$ &    1.4444 &    1.4439 &    0.0005 \,( 17) \\
$X$032 & $abcdaecdbe$ &    0.0055 &    0.0047 &    0.0008 \,(  8) \\
$X$033 & $abcdaedbce$ &   -0.3148 &   -0.3151 &    0.0003 \,(  7) \\
$X$034 & $abcdaedcbe$ &    0.7023 &    0.7042 &   -0.0019 \,(  9) \\
$X$035 & $abcdbeaced$ &    0.0509 &    0.0507 &    0.0002 \,(  3) \\
$X$036 & $abcdbecdea$ &    0.6736 &    0.6715 &    0.0021 \,( 22) \\
$X$037 & $abcdbedaec$ &   -0.4573 &   -0.4567 &   -0.0006 \,(  5) \\
$X$038 & $abcdbedcea$ &   -0.4278 &   -0.4330 &    0.0052 \,( 20) \\
$X$039 & $abcdceaebd$ &    0.7588 &    0.7610 &   -0.0021 \,( 13) \\
$X$040 & $abcdcebeda$ &    1.7368 &    1.7285 &    0.0083 \,( 52) \\
$X$041 & $abcdcedeab$ &    1.2820 &    1.2666 &    0.0154 \,( 77) \\
$X$042 & $abcdcedeba$ &   -0.8418 &   -0.8407 &   -0.0012 \,(121) \\
$X$043 & $abcddeeabc$ &   -2.4962 &   -2.4961 &   -0.0001 \,( 28) \\
$X$044 & $abcddeebca$ &   -2.2742 &   -2.2773 &    0.0031 \,(116) \\
$X$045 & $abcddeecab$ &    2.9694 &    2.9716 &   -0.0022 \,( 89) \\
$X$046 & $abcddeecba$ &    0.3265 &    0.3291 &   -0.0026 \,(117) \\
$X$047 & $abcdeabcde$ &   -3.1486 &   -3.1509 &    0.0023 \,( 21) \\
$X$048 & $abcdeacdbe$ &   -0.7549 &   -0.7553 &    0.0004 \,(  9) \\
$X$049 & $abcdeadbce$ &    0.0468 &    0.0469 &   -0.0001 \,(  7) \\
$X$050 & $abcdeadcbe$ &   -0.0469 &   -0.0483 &    0.0014 \,( 10) \\
$X$051 & $abcdebaced$ &   -0.3887 &   -0.3894 &    0.0008 \,(  6) \\
$X$052 & $abcdebcdea$ &   -3.2419 &   -3.2412 &   -0.0008 \,( 27) \\
$X$053 & $abcdebdaec$ &    0.4951 &    0.4955 &   -0.0004 \,(  5) \\
$X$054 & $abcdebdcea$ &    0.9817 &    0.9783 &    0.0034 \,( 15) \\
$X$055 & $abcdecaebd$ &   -0.1393 &   -0.1391 &   -0.0001 \,(  3) \\
$X$056 & $abcdecbeda$ &   -0.9680 &   -0.9725 &    0.0045 \,( 27) \\
$X$057 & $abcdecdeab$ &   -1.6133 &   -1.6098 &   -0.0035 \,( 32) \\
$X$058 & $abcdecdeba$ &    3.6409 &    3.6377 &    0.0032 \,( 67) \\
$X$059 & $abcdedeabc$ &    1.0735 &    1.0729 &    0.0005 \,( 28) \\
$X$060 & $abcdedebca$ &    3.9888 &    4.0064 &   -0.0176 \,(111) \\
$X$061 & $abcdedecab$ &   -3.8108 &   -3.8008 &   -0.0100 \,( 90) \\
$X$062 & $abcdedecba$ &   -0.9661 &   -0.9727 &    0.0066 \,(116) \\
$X$063 & $abcdeeabcd$ &    3.6700 &    3.6708 &   -0.0007 \,( 18) \\
$X$064 & $abcdeeacbd$ &   -0.6865 &   -0.6897 &    0.0032 \,( 14) \\
$X$065 & $abcdeebadc$ &   -0.3950 &   -0.3995 &    0.0045 \,( 16) \\
$X$066 & $abcdeebcda$ &    1.3654 &    1.3654 &    0.0000 \,( 43) \\
$X$067 & $abcdeecdab$ &    0.0999 &    0.1133 &   -0.0134 \,( 80) \\
$X$068 & $abcdeecdba$ &   -1.2657 &   -1.2820 &    0.0163 \,(116) \\
$X$069 & $abcdeedabc$ &    2.0979 &    2.0992 &   -0.0013 \,( 29) \\
$X$070 & $abcdeedbca$ &   -2.2167 &   -2.2247 &    0.0080 \,( 91) \\
$X$071 & $abcdeedcab$ &    1.7250 &    1.7292 &   -0.0042 \,( 87) \\
$X$072 & $abcdeedcba$ &    1.0015 &    1.0067 &   -0.0052 \,(111) \\
$X$073 & $abacbdceed$ &    1.0008 &    0.9701 &    0.0306 \,(137) \\
$X$074 & $abacbddece$ &   -2.2536 &   -2.2590 &    0.0054 \,(146) \\
$X$075 & $abacbddeec$ &    4.8166 &    4.8428 &   -0.0262 \,(157) \\
$X$076 & $abacbdecde$ &   -1.9252 &   -1.9264 &    0.0012 \,( 47) \\
$X$077 & $abacbdeced$ &    1.6928 &    1.7005 &   -0.0077 \,( 72) \\
$X$078 & $abacbdedce$ &   -1.0554 &   -1.0507 &   -0.0047 \,( 71) \\
$X$079 & $abacbdedec$ &   -9.0618 &   -9.0830 &    0.0211 \,(149) \\
$X$080 & $abacbdeecd$ &    2.7746 &    2.7441 &    0.0304 \,(130) \\
$X$081 & $abacbdeedc$ &    0.9199 &    0.9241 &   -0.0042 \,(138) \\
$X$082 & $abaccdbeed$ &    9.2505 &    9.2667 &   -0.0162 \,(156) \\
$X$083 & $abaccddeeb$ &   -8.4872 &   -8.4958 &    0.0085 \,(180) \\
$X$084 & $abaccdebde$ &   -3.3008 &   -3.2963 &   -0.0045 \,(117) \\
$X$085 & $abaccdebed$ &   -7.2082 &   -7.2344 &    0.0262 \,(128) \\
$X$086 & $abaccdedbe$ &   -2.7855 &   -2.7963 &    0.0108 \,(131) \\
$X$087 & $abaccdedeb$ &   13.2266 &   13.2390 &   -0.0124 \,(159) \\
$X$088 & $abaccdeebd$ &    6.2237 &    6.2243 &   -0.0006 \,(139) \\
$X$089 & $abaccdeedb$ &   -6.2947 &   -6.2968 &    0.0021 \,(147) \\
$X$090 & $abacdbceed$ &   -3.2459 &   -3.2783 &    0.0324 \,(115) \\
$X$091 & $abacdbdece$ &   -2.6500 &   -2.6397 &   -0.0102 \,( 79) \\
$X$092 & $abacdbdeec$ &   -2.0581 &   -2.0740 &    0.0159 \,(132) \\
$X$093 & $abacdbecde$ &   -0.1290 &   -0.1283 &   -0.0007 \,( 30) \\
$X$094 & $abacdbeced$ &    3.1484 &    3.1480 &    0.0004 \,( 34) \\
$X$095 & $abacdbedce$ &    0.4615 &    0.4633 &   -0.0018 \,( 20) \\
$X$096 & $abacdbedec$ &    1.3369 &    1.3310 &    0.0059 \,( 50) \\
$X$097 & $abacdbeecd$ &   -0.7685 &   -0.7749 &    0.0064 \,( 85) \\
$X$098 & $abacdbeedc$ &   -4.8059 &   -4.8299 &    0.0240 \,(103) \\
$X$099 & $abacdcbeed$ &    0.8280 &    0.8121 &    0.0159 \,(125) \\
$X$100 & $abacdcdeeb$ &    8.7993 &    8.7792 &    0.0201 \,(161) \\
$X$101 & $abacdcebde$ &    3.4269 &    3.4250 &    0.0019 \,( 35) \\
$X$102 & $abacdcebed$ &   -3.7779 &   -3.7771 &   -0.0008 \,( 58) \\
$X$103 & $abacdcedbe$ &   -1.4241 &   -1.4196 &   -0.0045 \,( 48) \\
$X$104 & $abacdcedeb$ &   -4.7053 &   -4.7163 &    0.0110 \,(150) \\
$X$105 & $abacdceebd$ &    0.9061 &    0.8914 &    0.0148 \,(114) \\
$X$106 & $abacdceedb$ &    2.3206 &    2.3088 &    0.0118 \,(150) \\
$X$107 & $abacddbeec$ &    3.3136 &    3.3079 &    0.0056 \,(137) \\
$X$108 & $abacddceeb$ &   -1.4732 &   -1.4651 &   -0.0080 \,(146) \\
$X$109 & $abacddebce$ &   -3.0206 &   -3.0355 &    0.0150 \,( 81) \\
$X$110 & $abacddebec$ &   -4.0784 &   -4.1024 &    0.0240 \,(121) \\
$X$111 & $abacddecbe$ &   -0.0006 &   -0.0170 &    0.0163 \,(116) \\
$X$112 & $abacddeceb$ &    2.8006 &    2.7898 &    0.0109 \,(145) \\
$X$113 & $abacddeebc$ &    6.2736 &    6.2572 &    0.0164 \,(126) \\
$X$114 & $abacddeecb$ &    0.5339 &    0.5109 &    0.0230 \,(143) \\
$X$115 & $abacdebcde$ &    2.7061 &    2.7062 &   -0.0001 \,( 33) \\
$X$116 & $abacdebced$ &   -1.0717 &   -1.0699 &   -0.0018 \,( 22) \\
$X$117 & $abacdebdce$ &   -1.3670 &   -1.3668 &   -0.0002 \,( 19) \\
$X$118 & $abacdebdec$ &    1.1957 &    1.1938 &    0.0019 \,( 32) \\
$X$119 & $abacdebecd$ &    2.2064 &    2.2074 &   -0.0010 \,( 28) \\
$X$120 & $abacdebedc$ &    1.4115 &    1.4109 &    0.0006 \,( 42) \\
$X$121 & $abacdecbde$ &   -1.6458 &   -1.6403 &   -0.0055 \,( 26) \\
$X$122 & $abacdecbed$ &    2.8738 &    2.8754 &   -0.0017 \,( 22) \\
$X$123 & $abacdecdbe$ &    0.4374 &    0.4341 &    0.0032 \,( 35) \\
$X$124 & $abacdecdeb$ &   -3.4547 &   -3.4485 &   -0.0062 \,(103) \\
$X$125 & $abacdecebd$ &   -2.1781 &   -2.1787 &    0.0005 \,( 44) \\
$X$126 & $abacdecedb$ &   -1.4862 &   -1.5082 &    0.0220 \,(117) \\
$X$127 & $abacdedbce$ &    3.3705 &    3.3752 &   -0.0047 \,( 29) \\
$X$128 & $abacdedbec$ &   -0.5696 &   -0.5721 &    0.0025 \,( 42) \\
$X$129 & $abacdedcbe$ &   -1.2506 &   -1.2450 &   -0.0056 \,( 43) \\
$X$130 & $abacdedceb$ &   -1.4870 &   -1.5081 &    0.0211 \,(117) \\
$X$131 & $abacdedebc$ &   -9.8525 &   -9.8687 &    0.0162 \,(125) \\
$X$132 & $abacdedecb$ &    0.1424 &    0.1562 &   -0.0139 \,(132) \\
$X$133 & $abacdeebcd$ &   -0.4303 &   -0.4381 &    0.0078 \,( 83) \\
$X$134 & $abacdeebdc$ &   -1.3482 &   -1.3680 &    0.0198 \,(114) \\
$X$135 & $abacdeecbd$ &    3.5080 &    3.4927 &    0.0153 \,( 82) \\
$X$136 & $abacdeecdb$ &    0.9296 &    0.9521 &   -0.0225 \,(129) \\
$X$137 & $abacdeedbc$ &    2.9713 &    2.9871 &   -0.0158 \,(122) \\
$X$138 & $abacdeedcb$ &    0.3293 &    0.3254 &    0.0039 \,(129) \\
$X$139 & $abbcaddeec$ &   -9.7784 &   -9.7779 &   -0.0005 \,(168) \\
$X$140 & $abbcadeced$ &   -5.1338 &   -5.1584 &    0.0246 \,(129) \\
$X$141 & $abbcadedec$ &   12.0790 &   12.1035 &   -0.0245 \,(155) \\
$X$142 & $abbcadeecd$ &    4.3702 &    4.3809 &   -0.0106 \,(135) \\
$X$143 & $abbcadeedc$ &   -4.2696 &   -4.2798 &    0.0101 \,(139) \\
$X$144 & $abbccdedea$ &   -4.5346 &   -4.5433 &    0.0087 \,(165) \\
$X$145 & $abbccdeeda$ &    1.3898 &    1.3889 &    0.0009 \,(165) \\
$X$146 & $abbcdadeec$ &    4.9932 &    5.0024 &   -0.0092 \,(137) \\
$X$147 & $abbcdaeced$ &   -1.7831 &   -1.7886 &    0.0055 \,( 55) \\
$X$148 & $abbcdaedec$ &    1.0690 &    1.0479 &    0.0211 \,( 80) \\
$X$149 & $abbcdaeecd$ &   -3.0595 &   -3.0547 &   -0.0048 \,(118) \\
$X$150 & $abbcdaeedc$ &    5.9109 &    5.9188 &   -0.0079 \,(113) \\
$X$151 & $abbcdcedea$ &    4.1124 &    4.1202 &   -0.0078 \,(153) \\
$X$152 & $abbcdceeda$ &   -4.9500 &   -4.9692 &    0.0192 \,(152) \\
$X$153 & $abbcddecea$ &   -5.1471 &   -5.1195 &   -0.0276 \,(155) \\
$X$154 & $abbcddeeca$ &   -2.0697 &   -2.0624 &   -0.0073 \,(142) \\
$X$155 & $abbcdeadec$ &   -0.4994 &   -0.5044 &    0.0050 \,( 49) \\
$X$156 & $abbcdeaedc$ &    0.0505 &    0.0464 &    0.0041 \,( 72) \\
$X$157 & $abbcdecdea$ &    5.7249 &    5.7098 &    0.0152 \,(120) \\
$X$158 & $abbcdeceda$ &    0.4457 &    0.4692 &   -0.0234 \,(119) \\
$X$159 & $abbcdedcea$ &    0.4635 &    0.4691 &   -0.0056 \,(127) \\
$X$160 & $abbcdedeca$ &   -4.5847 &   -4.5872 &    0.0025 \,(135) \\
$X$161 & $abbcdeecda$ &   -4.2537 &   -4.2675 &    0.0137 \,(132) \\
$X$162 & $abbcdeedca$ &    4.0560 &    4.0592 &   -0.0031 \,(127) \\
$X$163 & $abcabdceed$ &   -2.0575 &   -2.0320 &   -0.0254 \,( 94) \\
$X$164 & $abcabddeec$ &   -2.6442 &   -2.6524 &    0.0081 \,(103) \\
$X$165 & $abcabdeced$ &   -0.1194 &   -0.1180 &   -0.0013 \,( 31) \\
$X$166 & $abcabdedce$ &   -0.5847 &   -0.5871 &    0.0024 \,( 33) \\
$X$167 & $abcabdedec$ &    1.0077 &    1.0058 &    0.0018 \,( 98) \\
$X$168 & $abcabdeecd$ &   -2.6190 &   -2.6106 &   -0.0084 \,(106) \\
$X$169 & $abcabdeedc$ &    3.8627 &    3.8629 &   -0.0002 \,(101) \\
$X$170 & $abcacdbeed$ &    1.8551 &    1.8440 &    0.0111 \,(129) \\
$X$171 & $abcacddeeb$ &    2.9018 &    2.9093 &   -0.0075 \,(138) \\
$X$172 & $abcacdebed$ &    0.5603 &    0.5657 &   -0.0054 \,( 50) \\
$X$173 & $abcacdedeb$ &   -5.1930 &   -5.2023 &    0.0092 \,(129) \\
$X$174 & $abcacdeebd$ &    2.8772 &    2.8546 &    0.0226 \,(112) \\
$X$175 & $abcacdeedb$ &   -1.7167 &   -1.7008 &   -0.0158 \,(109) \\
$X$176 & $abcadbceed$ &   -1.0171 &   -1.0246 &    0.0075 \,( 41) \\
$X$177 & $abcadbdeec$ &   -2.6128 &   -2.6275 &    0.0146 \,( 79) \\
$X$178 & $abcadbeced$ &    0.8511 &    0.8515 &   -0.0004 \,( 20) \\
$X$179 & $abcadbedce$ &    0.3615 &    0.3623 &   -0.0008 \,(  7) \\
$X$180 & $abcadbedec$ &   -2.1100 &   -2.1099 &   -0.0002 \,( 18) \\
$X$181 & $abcadbeecd$ &   -4.9382 &   -4.9484 &    0.0102 \,( 30) \\
$X$182 & $abcadbeedc$ &   -1.7039 &   -1.7077 &    0.0038 \,( 36) \\
$X$183 & $abcadcbeed$ &   -0.9227 &   -0.9267 &    0.0040 \,( 32) \\
$X$184 & $abcadcdeeb$ &    0.8184 &    0.7872 &    0.0312 \,(122) \\
$X$185 & $abcadcebed$ &    2.7180 &    2.7170 &    0.0011 \,( 15) \\
$X$186 & $abcadcedeb$ &    0.0838 &    0.0854 &   -0.0016 \,( 26) \\
$X$187 & $abcadceebd$ &   -2.9121 &   -2.9123 &    0.0001 \,( 27) \\
$X$188 & $abcadceedb$ &    2.4254 &    2.4219 &    0.0035 \,( 41) \\
$X$189 & $abcaddbeec$ &    0.1034 &    0.1124 &   -0.0090 \,( 94) \\
$X$190 & $abcaddceeb$ &    2.3616 &    2.3607 &    0.0010 \,(109) \\
$X$191 & $abcaddebec$ &   -0.9936 &   -1.0026 &    0.0090 \,( 50) \\
$X$192 & $abcaddeceb$ &    2.1291 &    2.1059 &    0.0233 \,( 87) \\
$X$193 & $abcaddeebc$ &   -2.1456 &   -2.1438 &   -0.0017 \,( 73) \\
$X$194 & $abcaddeecb$ &    3.5787 &    3.5850 &   -0.0062 \,( 97) \\
$X$195 & $abcadebcde$ &   -1.0907 &   -1.0893 &   -0.0014 \,( 15) \\
$X$196 & $abcadebced$ &   -0.1423 &   -0.1447 &    0.0025 \,( 14) \\
$X$197 & $abcadebdce$ &    0.2431 &    0.2440 &   -0.0008 \,(  6) \\
$X$198 & $abcadebdec$ &    1.1699 &    1.1681 &    0.0018 \,( 15) \\
$X$199 & $abcadebecd$ &    0.9229 &    0.9249 &   -0.0020 \,( 19) \\
$X$200 & $abcadebedc$ &   -1.1453 &   -1.1476 &    0.0023 \,( 20) \\
$X$201 & $abcadecbde$ &   -0.9161 &   -0.9166 &    0.0005 \,( 12) \\
$X$202 & $abcadecbed$ &    1.4461 &    1.4499 &   -0.0039 \,( 11) \\
$X$203 & $abcadecdbe$ &    0.7071 &    0.7089 &   -0.0018 \,(  9) \\
$X$204 & $abcadecdeb$ &   -1.7229 &   -1.7237 &    0.0008 \,( 19) \\
$X$205 & $abcadecebd$ &    0.9296 &    0.9304 &   -0.0008 \,( 15) \\
$X$206 & $abcadecedb$ &    0.4456 &    0.4493 &   -0.0037 \,( 22) \\
$X$207 & $abcadedbce$ &    0.2029 &    0.2050 &   -0.0021 \,( 18) \\
$X$208 & $abcadedbec$ &   -0.9559 &   -0.9568 &    0.0009 \,( 16) \\
$X$209 & $abcadedcbe$ &    0.3514 &    0.3530 &   -0.0016 \,( 13) \\
$X$210 & $abcadedceb$ &   -0.5021 &   -0.5032 &    0.0012 \,( 21) \\
$X$211 & $abcadedebc$ &    4.6175 &    4.6186 &   -0.0011 \,( 77) \\
$X$212 & $abcadedecb$ &   -5.3537 &   -5.3696 &    0.0158 \,( 74) \\
$X$213 & $abcadeebcd$ &   -2.3729 &   -2.3817 &    0.0088 \,( 29) \\
$X$214 & $abcadeebdc$ &   -2.1724 &   -2.1755 &    0.0030 \,( 38) \\
$X$215 & $abcadeecbd$ &   -0.8326 &   -0.8377 &    0.0051 \,( 27) \\
$X$216 & $abcadeecdb$ &    1.5711 &    1.5544 &    0.0167 \,( 40) \\
$X$217 & $abcadeedbc$ &    2.0783 &    2.0828 &   -0.0045 \,( 64) \\
$X$218 & $abcadeedcb$ &   -0.9383 &   -0.9311 &   -0.0072 \,( 67) \\
$X$219 & $abcbaddeec$ &    7.9880 &    8.0015 &   -0.0135 \,(130) \\
$X$220 & $abcbadedec$ &   -7.7735 &   -7.7902 &    0.0166 \,(127) \\
$X$221 & $abcbadeecd$ &   -2.2739 &   -2.2953 &    0.0214 \,(108) \\
$X$222 & $abcbadeedc$ &    1.5524 &    1.5752 &   -0.0228 \,(107) \\
$X$223 & $abcbcdeeda$ &   -5.5966 &   -5.6242 &    0.0276 \,(140) \\
$X$224 & $abcbdadeec$ &    2.2881 &    2.2806 &    0.0075 \,(110) \\
$X$225 & $abcbdaedec$ &   -0.8721 &   -0.8731 &    0.0010 \,( 35) \\
$X$226 & $abcbdaeecd$ &   -0.5613 &   -0.5671 &    0.0058 \,( 50) \\
$X$227 & $abcbdaeedc$ &    1.6399 &    1.6057 &    0.0341 \,( 94) \\
$X$228 & $abcbdceeda$ &    5.6191 &    5.6524 &   -0.0333 \,(144) \\
$X$229 & $abcbddaeec$ &    2.6200 &    2.6301 &   -0.0101 \,(124) \\
$X$230 & $abcbddeeca$ &  -12.1174 &  -12.1359 &    0.0185 \,(155) \\
$X$231 & $abcbdeadec$ &    0.1871 &    0.1872 &   -0.0001 \,( 20) \\
$X$232 & $abcbdeaedc$ &   -1.2561 &   -1.2584 &    0.0023 \,( 27) \\
$X$233 & $abcbdecdea$ &   -8.9630 &   -8.9524 &   -0.0105 \,(108) \\
$X$234 & $abcbdeceda$ &    0.8243 &    0.7925 &    0.0318 \,(125) \\
$X$235 & $abcbdedaec$ &   -0.7272 &   -0.7295 &    0.0024 \,( 32) \\
$X$236 & $abcbdedcea$ &    1.0863 &    1.0565 &    0.0297 \,(127) \\
$X$237 & $abcbdedeca$ &   15.5666 &   15.5613 &    0.0053 \,(143) \\
$X$238 & $abcbdeeadc$ &    3.8779 &    3.8747 &    0.0033 \,( 54) \\
$X$239 & $abcbdeecda$ &    4.9499 &    4.9670 &   -0.0171 \,(133) \\
$X$240 & $abcbdeedca$ &  -10.4524 &  -10.4434 &   -0.0090 \,(130) \\
$X$241 & $abccaddeeb$ &    0.9042 &    0.8964 &    0.0078 \,(146) \\
$X$242 & $abccadedeb$ &   -0.4659 &   -0.4577 &   -0.0082 \,(137) \\
$X$243 & $abccadeedb$ &   -1.7181 &   -1.7289 &    0.0107 \,(130) \\
$X$244 & $abccdadeeb$ &    0.7519 &    0.7538 &   -0.0019 \,(131) \\
$X$245 & $abccdaedeb$ &    1.8008 &    1.7727 &    0.0281 \,( 83) \\
$X$246 & $abccdaeedb$ &    2.7987 &    2.8243 &   -0.0257 \,(111) \\
$X$247 & $abccddaeeb$ &    5.0799 &    5.0749 &    0.0050 \,(129) \\
$X$248 & $abccddeaeb$ &    3.4721 &    3.4735 &   -0.0014 \,(114) \\
$X$249 & $abccdeadeb$ &   -0.3379 &   -0.3440 &    0.0061 \,( 43) \\
$X$250 & $abccdeaedb$ &    1.3536 &    1.3451 &    0.0084 \,( 66) \\
$X$251 & $abccdedaeb$ &    0.6311 &    0.6121 &    0.0189 \,( 77) \\
$X$252 & $abccdedeab$ &   -2.7074 &   -2.6904 &   -0.0170 \,(115) \\
$X$253 & $abccdedeba$ &   -4.6305 &   -4.6586 &    0.0281 \,(142) \\
$X$254 & $abccdeeadb$ &    1.9926 &    2.0106 &   -0.0180 \,( 94) \\
$X$255 & $abccdeedab$ &    3.9969 &    3.9887 &    0.0082 \,(113) \\
$X$256 & $abccdeedba$ &    1.9506 &    1.9598 &   -0.0092 \,(132) \\
$X$257 & $abcdabceed$ &    1.6470 &    1.6547 &   -0.0077 \,( 51) \\
$X$258 & $abcdabdeec$ &    0.1470 &    0.1332 &    0.0138 \,( 39) \\
$X$259 & $abcdabeced$ &    0.9833 &    0.9816 &    0.0016 \,( 23) \\
$X$260 & $abcdabedec$ &   -0.9128 &   -0.9113 &   -0.0015 \,( 16) \\
$X$261 & $abcdabeecd$ &    5.8811 &    5.8858 &   -0.0047 \,( 31) \\
$X$262 & $abcdabeedc$ &   -1.3467 &   -1.3526 &    0.0059 \,( 25) \\
$X$263 & $abcdacbeed$ &   -0.1973 &   -0.2097 &    0.0124 \,( 35) \\
$X$264 & $abcdacdeeb$ &   -0.2238 &   -0.2310 &    0.0072 \,( 62) \\
$X$265 & $abcdacebed$ &   -2.0922 &   -2.0929 &    0.0007 \,( 14) \\
$X$266 & $abcdacedeb$ &    0.2385 &    0.2399 &   -0.0014 \,( 17) \\
$X$267 & $abcdaceebd$ &   -0.4459 &   -0.4509 &    0.0050 \,( 19) \\
$X$268 & $abcdaceedb$ &   -0.0754 &   -0.0849 &    0.0095 \,( 31) \\
$X$269 & $abcdadbeec$ &   -3.9083 &   -3.9166 &    0.0084 \,( 53) \\
$X$270 & $abcdadceeb$ &    0.7809 &    0.7592 &    0.0216 \,( 96) \\
$X$271 & $abcdadebec$ &    2.0365 &    2.0358 &    0.0006 \,( 21) \\
$X$272 & $abcdadeceb$ &   -2.4722 &   -2.4699 &   -0.0023 \,( 28) \\
$X$273 & $abcdadeebc$ &   -2.3810 &   -2.3912 &    0.0101 \,( 43) \\
$X$274 & $abcdadeecb$ &   -0.2085 &   -0.2252 &    0.0167 \,( 67) \\
$X$275 & $abcdaebced$ &   -0.6414 &   -0.6389 &   -0.0026 \,( 12) \\
$X$276 & $abcdaebdce$ &   -0.6241 &   -0.6252 &    0.0010 \,(  9) \\
$X$277 & $abcdaebdec$ &    1.9253 &    1.9273 &   -0.0021 \,(  9) \\
$X$278 & $abcdaebecd$ &    0.7919 &    0.7905 &    0.0015 \,( 21) \\
$X$279 & $abcdaebedc$ &   -0.2689 &   -0.2666 &   -0.0023 \,( 16) \\
$X$280 & $abcdaecbed$ &   -0.7398 &   -0.7391 &   -0.0007 \,(  8) \\
$X$281 & $abcdaecdeb$ &   -0.3317 &   -0.3322 &    0.0005 \,( 18) \\
$X$282 & $abcdaecebd$ &   -1.3421 &   -1.3398 &   -0.0022 \,( 15) \\
$X$283 & $abcdaecedb$ &    0.6066 &    0.6088 &   -0.0022 \,( 19) \\
$X$284 & $abcdaedbec$ &   -0.2388 &   -0.2385 &   -0.0003 \,(  7) \\
$X$285 & $abcdaedceb$ &    0.6416 &    0.6419 &   -0.0003 \,( 14) \\
$X$286 & $abcdaedebc$ &   -1.0218 &   -1.0202 &   -0.0015 \,( 17) \\
$X$287 & $abcdaedecb$ &    0.3736 &    0.3707 &    0.0029 \,( 24) \\
$X$288 & $abcdaeebcd$ &    3.1220 &    3.1237 &   -0.0017 \,( 31) \\
$X$289 & $abcdaeebdc$ &   -1.8638 &   -1.8697 &    0.0060 \,( 24) \\
$X$290 & $abcdaeecbd$ &   -0.7830 &   -0.7868 &    0.0038 \,( 24) \\
$X$291 & $abcdaeecdb$ &    0.4308 &    0.4271 &    0.0037 \,( 33) \\
$X$292 & $abcdaeedbc$ &   -2.4709 &   -2.4746 &    0.0037 \,( 34) \\
$X$293 & $abcdaeedcb$ &    0.7900 &    0.7841 &    0.0058 \,( 47) \\
$X$294 & $abcdbaceed$ &   -1.2498 &   -1.2586 &    0.0088 \,( 28) \\
$X$295 & $abcdbadeec$ &   -1.6649 &   -1.6653 &    0.0004 \,( 29) \\
$X$296 & $abcdbaeced$ &   -0.1647 &   -0.1627 &   -0.0020 \,( 15) \\
$X$297 & $abcdbaedec$ &    0.7932 &    0.7933 &   -0.0001 \,( 15) \\
$X$298 & $abcdbaeecd$ &   -2.0631 &   -2.0751 &    0.0119 \,( 27) \\
$X$299 & $abcdbaeedc$ &   -2.1978 &   -2.2016 &    0.0038 \,( 24) \\
$X$300 & $abcdbceeda$ &    6.9313 &    6.9188 &    0.0125 \,(120) \\
$X$301 & $abcdbdaeec$ &    1.9082 &    1.9004 &    0.0078 \,( 78) \\
$X$302 & $abcdbdeeca$ &    3.1366 &    3.1515 &   -0.0149 \,(123) \\
$X$303 & $abcdbeadec$ &    0.6740 &    0.6751 &   -0.0011 \,(  6) \\
$X$304 & $abcdbeaecd$ &   -1.1747 &   -1.1735 &   -0.0012 \,( 15) \\
$X$305 & $abcdbeaedc$ &    1.6599 &    1.6597 &    0.0002 \,( 13) \\
$X$306 & $abcdbeceda$ &   -7.3924 &   -7.3953 &    0.0028 \,( 54) \\
$X$307 & $abcdbedeca$ &    3.1216 &    3.1065 &    0.0150 \,( 65) \\
$X$308 & $abcdbeeadc$ &   -1.1535 &   -1.1575 &    0.0039 \,( 26) \\
$X$309 & $abcdbeecda$ &    5.8275 &    5.8320 &   -0.0045 \,( 98) \\
$X$310 & $abcdbeedca$ &    0.3623 &    0.3828 &   -0.0204 \,(112) \\
$X$311 & $abcdcabeed$ &   -2.5341 &   -2.5504 &    0.0162 \,( 52) \\
$X$312 & $abcdcadeeb$ &    1.0401 &    1.0207 &    0.0194 \,( 86) \\
$X$313 & $abcdcaebed$ &    2.1802 &    2.1831 &   -0.0028 \,( 22) \\
$X$314 & $abcdcaedeb$ &    0.4125 &    0.4170 &   -0.0045 \,( 28) \\
$X$315 & $abcdcaeebd$ &   -3.1553 &   -3.1676 &    0.0123 \,( 41) \\
$X$316 & $abcdcaeedb$ &   -0.3966 &   -0.4056 &    0.0090 \,( 46) \\
$X$317 & $abcdcbeeda$ &    1.7433 &    1.7588 &   -0.0155 \,(127) \\
$X$318 & $abcdcdaeeb$ &    3.9324 &    3.9430 &   -0.0107 \,(125) \\
$X$319 & $abcdcdeaeb$ &   -5.9699 &   -5.9938 &    0.0240 \,(116) \\
$X$320 & $abcdceadeb$ &    0.3668 &    0.3680 &   -0.0012 \,( 20) \\
$X$321 & $abcdceaedb$ &   -2.4587 &   -2.4577 &   -0.0010 \,( 25) \\
$X$322 & $abcdcedaeb$ &   -0.5372 &   -0.5373 &    0.0001 \,( 22) \\
$X$323 & $abcdceeadb$ &    2.9771 &    2.9709 &    0.0062 \,( 46) \\
$X$324 & $abcdceedab$ &   -3.5381 &   -3.5156 &   -0.0225 \,(120) \\
$X$325 & $abcdceedba$ &   -1.3839 &   -1.4022 &    0.0183 \,(136) \\
$X$326 & $abcddabeec$ &   -5.0526 &   -5.0717 &    0.0192 \,( 95) \\
$X$327 & $abcddaceeb$ &    1.2461 &    1.2610 &   -0.0149 \,(107) \\
$X$328 & $abcddaebec$ &   -1.7301 &   -1.7381 &    0.0080 \,( 41) \\
$X$329 & $abcddaeceb$ &    1.3418 &    1.3143 &    0.0275 \,( 64) \\
$X$330 & $abcddaeebc$ &   -3.8303 &   -3.8477 &    0.0175 \,( 74) \\
$X$331 & $abcddaeecb$ &    4.3816 &    4.4105 &   -0.0289 \,(113) \\
$X$332 & $abcddbaeec$ &    3.3710 &    3.4038 &   -0.0328 \,(112) \\
$X$333 & $abcddbeeca$ &   -7.3732 &   -7.3712 &   -0.0020 \,(134) \\
$X$334 & $abcddcaeeb$ &   -2.0019 &   -2.0111 &    0.0092 \,(120) \\
$X$335 & $abcddceaeb$ &   -2.0635 &   -2.0346 &   -0.0289 \,(116) \\
$X$336 & $abcddeabec$ &   -0.7452 &   -0.7516 &    0.0063 \,( 22) \\
$X$337 & $abcddeaceb$ &   -0.3840 &   -0.3922 &    0.0082 \,( 32) \\
$X$338 & $abcddeaebc$ &   -1.7560 &   -1.7643 &    0.0083 \,( 38) \\
$X$339 & $abcddeaecb$ &    1.3648 &    1.3668 &   -0.0020 \,( 43) \\
$X$340 & $abcddebeca$ &    2.7835 &    2.7951 &   -0.0116 \,(116) \\
$X$341 & $abcddecaeb$ &    1.4617 &    1.4569 &    0.0048 \,( 37) \\
$X$342 & $abcddeeacb$ &    0.6581 &    0.6621 &   -0.0040 \,( 91) \\
$X$343 & $abcdeabced$ &    2.6108 &    2.6150 &   -0.0041 \,( 17) \\
$X$344 & $abcdeabdce$ &    2.0240 &    2.0275 &   -0.0035 \,( 13) \\
$X$345 & $abcdeabdec$ &   -1.2875 &   -1.2859 &   -0.0016 \,( 16) \\
$X$346 & $abcdeabecd$ &   -0.0366 &   -0.0333 &   -0.0034 \,( 10) \\
$X$347 & $abcdeabedc$ &   -0.2194 &   -0.2215 &    0.0020 \,( 12) \\
$X$348 & $abcdeacbed$ &   -0.1230 &   -0.1247 &    0.0018 \,( 11) \\
$X$349 & $abcdeacdeb$ &    1.2189 &    1.2209 &   -0.0020 \,( 23) \\
$X$350 & $abcdeacebd$ &    1.1395 &    1.1305 &    0.0090 \,(  8) \\
$X$351 & $abcdeacedb$ &   -0.6324 &   -0.6323 &   -0.0000 \,( 13) \\
$X$352 & $abcdeadbec$ &    0.5207 &    0.5105 &    0.0103 \,(  6) \\
$X$353 & $abcdeadceb$ &   -0.3938 &   -0.3928 &   -0.0011 \,( 13) \\
$X$354 & $abcdeadebc$ &    1.5958 &    1.5966 &   -0.0007 \,( 14) \\
$X$355 & $abcdeadecb$ &   -0.9731 &   -0.9744 &    0.0013 \,( 15) \\
$X$356 & $abcdeaebcd$ &    1.8414 &    1.8419 &   -0.0006 \,( 22) \\
$X$357 & $abcdeaebdc$ &   -0.5362 &   -0.5349 &   -0.0014 \,( 15) \\
$X$358 & $abcdeaecbd$ &   -1.1596 &   -1.1584 &   -0.0012 \,( 15) \\
$X$359 & $abcdeaecdb$ &    0.6688 &    0.6679 &    0.0009 \,( 16) \\
$X$360 & $abcdeaedbc$ &   -1.2252 &   -1.2258 &    0.0006 \,( 16) \\
$X$361 & $abcdeaedcb$ &    0.3427 &    0.3457 &   -0.0030 \,( 28) \\
$X$362 & $abcdebadec$ &   -0.6576 &   -0.6580 &    0.0004 \,( 13) \\
$X$363 & $abcdebaecd$ &   -0.7977 &   -0.8003 &    0.0027 \,( 11) \\
$X$364 & $abcdebaedc$ &    1.3965 &    1.3995 &   -0.0030 \,( 12) \\
$X$365 & $abcdebceda$ &    3.0356 &    3.0311 &    0.0045 \,( 33) \\
$X$366 & $abcdebdeca$ &   -2.2969 &   -2.2983 &    0.0015 \,( 48) \\
$X$367 & $abcdebeadc$ &    0.8945 &    0.8931 &    0.0014 \,( 15) \\
$X$368 & $abcdebecda$ &   -5.7003 &   -5.7062 &    0.0059 \,( 45) \\
$X$369 & $abcdebedca$ &    2.0779 &    2.0753 &    0.0026 \,( 63) \\
$X$370 & $abcdecadeb$ &   -0.1662 &   -0.1676 &    0.0014 \,( 15) \\
$X$371 & $abcdecaedb$ &    0.7278 &    0.7282 &   -0.0005 \,( 11) \\
$X$372 & $abcdecdaeb$ &   -1.3225 &   -1.3236 &    0.0011 \,( 18) \\
$X$373 & $abcdeceadb$ &   -0.5533 &   -0.5527 &   -0.0007 \,( 20) \\
$X$374 & $abcdecedab$ &    1.6342 &    1.6264 &    0.0078 \,( 98) \\
$X$375 & $abcdecedba$ &   -3.3518 &   -3.3366 &   -0.0152 \,( 98) \\
$X$376 & $abcdedabec$ &   -0.5170 &   -0.5120 &   -0.0051 \,( 15) \\
$X$377 & $abcdedaceb$ &    0.7619 &    0.7627 &   -0.0008 \,( 17) \\
$X$378 & $abcdedaebc$ &   -1.0546 &   -1.0533 &   -0.0013 \,( 16) \\
$X$379 & $abcdedaecb$ &    0.1342 &    0.1352 &   -0.0010 \,( 23) \\
$X$380 & $abcdedbeca$ &    3.2248 &    3.2321 &   -0.0073 \,( 55) \\
$X$381 & $abcdedcaeb$ &    0.1205 &    0.1226 &   -0.0021 \,( 22) \\
$X$382 & $abcdedeacb$ &    0.2570 &    0.2188 &    0.0382 \,( 76) \\
$X$383 & $abcdeeabdc$ &   -2.9396 &   -2.9494 &    0.0097 \,( 24) \\
$X$384 & $abcdeeacdb$ &    1.0003 &    0.9920 &    0.0083 \,( 32) \\
$X$385 & $abcdeeadbc$ &   -3.3498 &   -3.3560 &    0.0062 \,( 22) \\
$X$386 & $abcdeeadcb$ &    1.9426 &    1.9312 &    0.0114 \,( 45) \\
$X$387 & $abcdeebdca$ &   -0.3093 &   -0.2943 &   -0.0150 \,( 82) \\
$X$388 & $abcdeecadb$ &    1.3253 &    1.3222 &    0.0031 \,( 36) \\
$X$389 & $abcdeedacb$ &   -2.6523 &   -2.6381 &   -0.0142 \,( 64) \\
\end{longtable*}
\endgroup

\renewcommand{\arraystretch}{1.0}

\begingroup
\squeezetable
\renewcommand{\arraystretch}{0.8}
\setlength\LTleft{0pt}
\setlength\LTright{0pt}
\setlength\LTcapwidth{\textwidth}
\begin{longtable*}{l@{\hskip-6em}d@{\hskip+2em}d@{\hskip+2em}d}
\caption*{%
\small TABLE~\ref{table:XB1B2}:
Numerical evaluation of the 98 integrals in Set~V.
Of the 389 integrals in Set~V, the 98 contain only a single second-order self-energy subdiagram and no other self-energy subdiagrams. The first column shows the diagram labels, the second and third columns display the old and new evaluations of the integrals, respectively, and the fourth column shows the difference between the values in the second and third columns.
\label{table:XB1B2}
} \\
\hline \hline
\multicolumn{1}{m{0.12\textwidth}}{Diagram $\mathcal{G}$}&
\multicolumn{1}{m{0.15\textwidth}}{Old Value}&
\multicolumn{1}{m{0.15\textwidth}}{New Value}&
\multicolumn{1}{m{0.12\textwidth}}{Difference}
\\
\hline
\endfirsthead
\caption*{%
\small TABLE~\ref{table:XB1B2} (continued):
Numerical evaluation of the 98 integrals in Set~V.
} \\
\hline \hline
\multicolumn{1}{m{0.12\textwidth}}{Diagram $\mathcal{G}$}&
\multicolumn{1}{m{0.15\textwidth}}{Old Value}&
\multicolumn{1}{m{0.15\textwidth}}{New Value}&
\multicolumn{1}{m{0.12\textwidth}}{Difference}
\\
\hline
\endhead
\hline
\endfoot
\hline \hline
\endlastfoot
$X$005  &  1.0993 \,( 94) &  1.0879 \,( 44) & -0.0114 \,(104) \\
$X$017  &  0.5151 \,( 11) &  0.5125 \,(  8) & -0.0026 \,( 14) \\
$X$018  &  0.0313 \,( 11) &  0.0286 \,(  9) & -0.0027 \,( 15) \\
$X$023  &  0.4498 \,( 49) &  0.4459 \,( 36) & -0.0039 \,( 61) \\
$X$063  &  3.4296 \,( 12) &  3.4299 \,( 10) &  0.0003 \,( 16) \\
$X$064  & -0.2773 \,(  7) & -0.2812 \,(  7) & -0.0038 \,( 10) \\
$X$065  &  0.1559 \,( 11) &  0.1539 \,(  8) & -0.0020 \,( 13) \\
$X$073  &  3.3860 \,(110) &  3.3704 \,( 62) & -0.0156 \,(127) \\
$X$074  &  4.3988 \,(113) &  4.4026 \,( 80) &  0.0038 \,(139) \\
$X$080  &  0.4507 \,(112) &  0.4276 \,( 43) & -0.0231 \,(120) \\
$X$084  &  8.9910 \,(110) &  8.9923 \,( 43) &  0.0012 \,(118) \\
$X$085  & -2.2901 \,(110) & -2.3162 \,( 45) & -0.0261 \,(118) \\
$X$086  &  0.5154 \,(111) &  0.5225 \,( 65) &  0.0071 \,(129) \\
$X$090  &  1.5228 \,(111) &  1.4899 \,( 43) & -0.0329 \,(119) \\
$X$092  &  2.1099 \,(114) &  2.1020 \,( 44) & -0.0079 \,(122) \\
$X$097  &  5.0296 \,( 81) &  5.0258 \,( 44) & -0.0038 \,( 92) \\
$X$098  & -2.0133 \,( 91) & -2.0332 \,( 44) & -0.0200 \,(101) \\
$X$099  &  3.0661 \,(106) &  3.0518 \,( 44) & -0.0143 \,(115) \\
$X$105  &  3.0334 \,(100) &  3.0212 \,( 44) & -0.0123 \,(109) \\
$X$109  & -0.0857 \,( 77) & -0.1017 \,( 45) & -0.0160 \,( 89) \\
$X$110  &  1.9054 \,(108) &  1.8855 \,( 45) & -0.0199 \,(117) \\
$X$111  &  3.3599 \,(108) &  3.3443 \,( 44) & -0.0156 \,(117) \\
$X$133  &  2.7279 \,( 79) &  2.7208 \,( 43) & -0.0071 \,( 90) \\
$X$134  & -0.6672 \,(104) & -0.6772 \,( 42) & -0.0100 \,(113) \\
$X$135  &  0.9086 \,( 70) &  0.8971 \,( 44) & -0.0116 \,( 83) \\
$X$140  & -2.8019 \,(113) & -2.8222 \,( 43) & -0.0202 \,(121) \\
$X$147  &  1.1259 \,( 48) &  1.1229 \,( 28) & -0.0031 \,( 56) \\
$X$148  & -1.4111 \,( 65) & -1.4227 \,( 44) & -0.0115 \,( 79) \\
$X$155  &  5.0349 \,( 39) &  5.0301 \,( 28) & -0.0048 \,( 48) \\
$X$156  & -0.8279 \,( 60) & -0.8249 \,( 44) &  0.0030 \,( 75) \\
$X$163  &  6.8241 \,( 83) &  6.8410 \,( 44) &  0.0169 \,( 94) \\
$X$168  &  3.4446 \,(102) &  3.4458 \,( 44) &  0.0012 \,(111) \\
$X$170  &  0.2295 \,(112) &  0.2097 \,( 44) & -0.0198 \,(121) \\
$X$174  &  1.6998 \,(102) &  1.6825 \,( 42) & -0.0173 \,(110) \\
$X$176  &  0.7446 \,( 30) &  0.7404 \,( 25) & -0.0042 \,( 40) \\
$X$177  &  0.0127 \,( 74) & -0.0009 \,( 34) & -0.0136 \,( 82) \\
$X$181  & -4.4353 \,( 25) & -4.4430 \,( 18) & -0.0077 \,( 31) \\
$X$182  &  1.2800 \,( 33) &  1.2756 \,( 16) & -0.0044 \,( 36) \\
$X$183  & -0.0790 \,( 25) & -0.0823 \,( 13) & -0.0032 \,( 29) \\
$X$184  &  0.2062 \,(113) &  0.1682 \,( 44) & -0.0380 \,(122) \\
$X$187  &  1.2699 \,( 23) &  1.2703 \,( 13) &  0.0004 \,( 26) \\
$X$188  &  1.7960 \,( 32) &  1.7910 \,( 18) & -0.0050 \,( 37) \\
$X$191  &  0.1892 \,( 46) &  0.1827 \,( 27) & -0.0065 \,( 53) \\
$X$192  &  2.3868 \,( 81) &  2.3657 \,( 27) & -0.0212 \,( 86) \\
$X$213  & -2.4516 \,( 23) & -2.4580 \,( 14) & -0.0065 \,( 28) \\
$X$214  &  0.6795 \,( 34) &  0.6750 \,( 14) & -0.0046 \,( 37) \\
$X$215  &  0.0734 \,( 21) &  0.0693 \,(  9) & -0.0040 \,( 23) \\
$X$216  & -1.3023 \,( 33) & -1.3182 \,( 17) & -0.0159 \,( 38) \\
$X$221  &  0.6703 \,( 99) &  0.6523 \,( 44) & -0.0179 \,(108) \\ 
$X$224  &  2.4794 \,( 96) &  2.4718 \,( 42) & -0.0076 \,(105) \\
$X$226  &  1.0424 \,( 46) &  1.0386 \,( 30) & -0.0038 \,( 55) \\
$X$227  &  0.5864 \,( 87) &  0.5530 \,( 30) & -0.0333 \,( 92) \\
$X$238  &  1.2765 \,( 41) &  1.2740 \,( 34) & -0.0025 \,( 53) \\
$X$245  &  0.0693 \,( 76) &  0.0515 \,( 44) & -0.0179 \,( 88) \\
$X$249  &  4.0125 \,( 37) &  4.0091 \,( 30) & -0.0034 \,( 48) \\
$X$250  & -1.0540 \,( 60) & -1.0579 \,( 44) & -0.0039 \,( 75) \\
$X$251  & -1.3846 \,( 71) & -1.3984 \,( 43) & -0.0138 \,( 83) \\
$X$257  &  5.7467 \,( 43) &  5.7495 \,( 32) &  0.0029 \,( 53) \\
$X$258  & -0.5252 \,( 32) & -0.5361 \,( 19) & -0.0109 \,( 37) \\
$X$261  &  6.4027 \,( 26) &  6.4077 \,( 29) &  0.0051 \,( 39) \\
$X$262  & -2.2852 \,( 20) & -2.2845 \,( 21) &  0.0008 \,( 29) \\
$X$263  & -2.8297 \,( 30) & -2.8392 \,( 14) & -0.0095 \,( 33) \\
$X$264  &  4.8803 \,( 53) &  4.8753 \,( 31) & -0.0050 \,( 62) \\
$X$267  & -0.6603 \,( 16) & -0.6643 \,( 12) & -0.0039 \,( 21) \\
$X$268  &  0.1192 \,( 27) &  0.1126 \,( 14) & -0.0066 \,( 31) \\
$X$269  & -0.7195 \,( 47) & -0.7249 \,( 29) & -0.0054 \,( 55) \\
$X$270  & -1.6851 \,( 88) & -1.6986 \,( 40) & -0.0135 \,( 97) \\
$X$273  & -2.0460 \,( 39) & -2.0517 \,( 33) & -0.0057 \,( 51) \\
$X$274  &  0.8701 \,( 61) &  0.8531 \,( 24) & -0.0170 \,( 66) \\
$X$288  &  4.2119 \,( 24) &  4.2106 \,( 23) & -0.0014 \,( 33) \\
$X$289  & -1.5654 \,( 17) & -1.5717 \,( 15) & -0.0063 \,( 23) \\
$X$290  & -3.7756 \,( 20) & -3.7779 \,( 12) & -0.0022 \,( 23) \\
$X$291  &  1.5943 \,( 28) &  1.5881 \,( 17) & -0.0062 \,( 33) \\
$X$292  &  0.9125 \,( 29) &  0.9090 \,( 13) & -0.0035 \,( 32) \\
$X$293  & -1.2629 \,( 33) & -1.2700 \,( 18) & -0.0071 \,( 38) \\
$X$294  & -3.3885 \,( 20) & -3.3930 \,( 18) & -0.0045 \,( 28) \\
$X$295  &  1.7891 \,( 23) &  1.7902 \,( 17) &  0.0011 \,( 29) \\
$X$298  & -1.9145 \,( 23) & -1.9212 \,( 13) & -0.0067 \,( 27) \\
$X$299  & -0.2902 \,( 19) & -0.2920 \,( 13) & -0.0018 \,( 23) \\
$X$301  & -1.3380 \,( 68) & -1.3514 \,( 40) & -0.0134 \,( 79) \\
$X$308  &  1.8123 \,( 19) &  1.8105 \,( 20) & -0.0018 \,( 28) \\
$X$311  & -0.5326 \,( 47) & -0.5430 \,( 31) & -0.0104 \,( 56) \\
$X$312  & -1.2622 \,( 76) & -1.2762 \,( 42) & -0.0141 \,( 87) \\
$X$315  & -1.3745 \,( 36) & -1.3837 \,( 25) & -0.0091 \,( 44) \\
$X$316  &  0.0123 \,( 36) &  0.0085 \,( 23) & -0.0037 \,( 43) \\
$X$323  & -0.0026 \,( 36) & -0.0099 \,( 22) & -0.0073 \,( 43) \\
$X$328  & -0.2772 \,( 36) & -0.2859 \,( 25) & -0.0086 \,( 44) \\
$X$329  & -0.9591 \,( 56) & -0.9788 \,( 26) & -0.0197 \,( 62) \\
$X$336  & -0.7700 \,( 17) & -0.7772 \,( 13) & -0.0073 \,( 21) \\
$X$337  & -1.2027 \,( 27) & -1.2088 \,( 20) & -0.0060 \,( 34) \\
$X$338  & -1.8502 \,( 33) & -1.8573 \,( 23) & -0.0071 \,( 41) \\
$X$339  &  0.4124 \,( 33) &  0.4130 \,( 26) &  0.0006 \,( 43) \\
$X$341  &  1.7808 \,( 30) &  1.7757 \,( 15) & -0.0051 \,( 33) \\
$X$383  & -4.8018 \,( 19) & -4.8082 \,( 15) & -0.0065 \,( 24) \\
$X$384  &  1.9293 \,( 27) &  1.9199 \,( 21) & -0.0094 \,( 35) \\
$X$385  & -0.7421 \,( 16) & -0.7436 \,( 13) & -0.0015 \,( 21) \\
$X$386  &  0.6884 \,( 32) &  0.6802 \,( 17) & -0.0083 \,( 37) \\
$X$388  & -0.4344 \,( 31) & -0.4474 \,( 14) & -0.0130 \,( 34) \\ 
\hline 
total & 57.8056  \,(638) & 57.0023 \,(327)& -0.8032 \,(717) \\
\end{longtable*}
\endgroup
\renewcommand{\arraystretch}{1.0}

}


\end{document}